\documentclass[epj]{webofc}
\usepackage[varg]{txfonts}   
\woctitle{International Conference on New Frontiers in Physics 2013}
\begin{document}
\title{Neutron Star masses from the Field Correlator Method Equation of State	
}

\author{D. Zappal\`a\inst{1}\fnsep\thanks{Speaker, \email{dario.zappala@ct.infn.it
   }} \and
        G. F. Burgio\inst{1}\fnsep\thanks{\email{fiorella.burgio@ct.infn.it
             }} \and
        V. Greco\inst{2,3}\fnsep\thanks{\email{greco@lns.infn.it
             }} \and
        S. Plumari\inst{2,3}\fnsep\thanks{\email{salvatore.plumari@ct.infn.it
             }}
}

\institute{INFN Sezione di Catania, Via Santa Sofia 64, I-95123 Catania, Italia
\and
           Dipartimento di Fisica e Astronomia, Universit\`a di Catania, Via Santa Sofia 64, I-95123 Catania, Italia
\and
          INFN - Laboratori Nazionali del Sud, Via Santa Sofia 62, I-95125 Catania, Italia 
          }

\abstract{%
We analyse the hadron-quark phase transition in neutron stars by confronting the hadronic Equation of State (EoS) 
obtained according to the microscopic Brueckner-Hartree-Fock  many body theory, with the quark matter EoS derived 
within the Field Correlator Method. In particular, the latter EoS is only parametrized in terms of 
the gluon condensate and the large distance quark-antiquark potential, so that the comparison of the 
results of this analysis with the most recent measurements of 
heavy neutron star masses provides some physical constraints on these two parameters.
}
\maketitle
\section{Introduction}
\label{intro}
The appearance of quark matter in the interior of massive neutron stars (NS) is one of the mostly
debated issues in the physics of these compact objects. 
If one considers only purely nucleonic degrees of freedom in the construction of 
the Equation of State (EoS) \cite{gabri} to describe the interior of NS,
it turns out that for the heaviest NS, close to the maximum mass (about two solar masses), 
the central particle density reaches values larger than $1/{\rm fm}^3$, so that 
the nucleon cores start to touch each other, and it is hard to imagine that only nucleonic 
degrees of freedom can play a role. On the contrary, it can be expected 
that even before reaching these density values, the nucleons start to lose their identity, 
and quark degrees of freedom are excited at a macroscopic level.

The value of the maximum mass of NS is probably one of the physical quantities that is most sensitive to the presence of quark matter in NS. 
The recent observation of a large NS mass in PSR J0348+0432 with mass $\rm M=2.01 \pm 0.04 M_\odot$ \cite{maxpuls} (see also  PSR J1614-2230 
with mass $\rm M=1.97 \pm 0.04 M_\odot$ \cite{maxpuls2})
implies that the EoS of NS matter is stiff enough to keep the maximum mass at these large values. 
While purely nucleonic EoS are able to accommodate such large masses  \cite{gabri}, 
the presence of non-nucleonic degrees of freedom, like hyperons and  quarks, tends usually to soften considerably the EoS, thus lowering the mass value, 
which could  
be incompatible with observations. The large value of the mass could then be explained only if  quark matter EoS are  particularly stiff. 
Unfortunately,  the quark matter EoS is poorly known at zero temperature and at the high baryonic density appropriate for NS. 
One has, therefore, to rely on models of quark matter and compare their predictions to estimate the uncertainty of the results 
for the NS matter as well as for the NS structure and mass. 

Here we report the main predictions, discussed in more detail in the two papers \cite{noi08,noi13},
which are obtained with two definite nucleonic EoS developed on the basis of 
the nonrelativistic  Brueckner-Hartree-Fock  (BHF) many-body theory for nuclear matter
and on its relativistic, Dirac-Brueckner-Hartree-Fock (DBHF), formulation,
and by adopting, for the quark EoS,  the  Field Correlator Model (FCM) \cite{phrep}
which in principle is able to cover the full temperature-chemical potential plane.
In particular, unlike the  Nambu--Jona-Lasinio model,  the FCM EoS contains {\it ab initio} the property of
confinement, which is expected to play an important role as far as the stability of a neutron star is concerned \cite{noi07}.
We shall focus on the predictions of the maximum NS mass by varying the two parameters of the FCM, namely the 
$\overline q~q$ potential $V_1$ and the gluon condensate $G_2$, also exploring the dependence 
of $G_2$  on the baryon chemical potential $\mu_B$.

In Sect.~\ref{sec-1} we briefly recall some features of the EoS both for nuclear matter and quark matter used in this analysis,
while in Sect.~\ref{results} details of the hadron-quark phase transition and related predictions for the NS masses are illustrated. 
The conclusions are reported in Sect.~\ref{conclusions}.

\begin{figure*}
\centering
\includegraphics[height=5cm,clip]{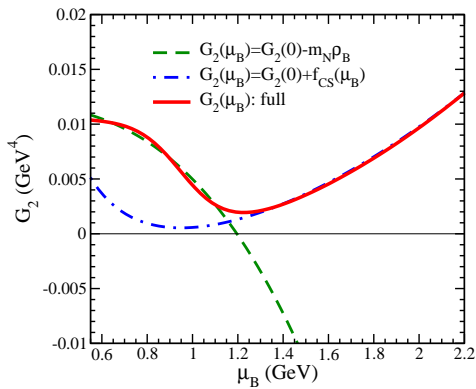}
\caption{$G_2(\mu_B)$ as computed in \cite{cohen} 
(dashed green) and in \cite{zit1} (dot-dashed blue) with $G_2(\mu_B= 0)= 0.012$ GeV$^4$.
The solid red line is the effective approximation used in our
analysis \cite{noi13}.
}
\label{fig1}   
\end{figure*}

\section{EoS for nuclear and quark matter}
\label{sec-1}
\subsection{Nuclear matter: the Brueckner-Bethe-Goldstone theory}
\label{sec-2}
Let us briefly recall the main features of the BHF method for the nuclear matter EoS.
This theoretical scheme is based on the Brueckner-Bethe-Goldstone  many-body theory, 
which is the linked cluster expansion of the energy per nucleon of nuclear matter 
(see Ref.\cite{book}, chapter 1 and references therein). 
In this approach the bare nucleon-nucleon interaction V is systematically replaced 
by the Brueckner reaction matrix G,  which  is the solution of the Bethe-Goldstone equation
\begin{equation}
G(\rho;\omega) =  V  + V \sum_{k_a k_b} \frac{|k_a k_b\rangle  Q  \langle k_a k_b|}
{\omega - e(k_a) - e(k_b) } G(\rho;\omega),
\end{equation}
\noindent
where $\rho$ is the nucleon number density, $\omega$  is the  starting energy, and
$|k_a k_b\rangle Q \langle k_a k_b|$  is  the Pauli operator.
$e(k;\rho) = k^2 / (2m) + U(k;\rho)$
is the single particle energy and $U(k;\rho)$ is the single particle potential.
In the BHF approximation, the energy per nucleon is (the subscript ``{\it a}'' indicates antisymmetrization of the
matrix element)
\begin{equation} 
\frac{E}{A}(\rho)  = \frac{3}{5} \frac{k_F^2}{2m} + \frac{1}{2A}
\sum_{k,k'\leq k_F} \langle k k'|G(\rho; e(k)+e(k'))|k k'\rangle_a \, .
\end{equation}
In this analysis the Argonne $v_{18}$ potential \cite{v18} is chosen as the nucleon-nucleon potential, 
supplemented  by the so-called Urbana model \cite{uix} as three-body force.
In fact, it is well known that the non-relativistic calculations, based on 
purely two-body interactions, fail to reproduce the correct saturation point of symmetric 
nuclear matter and one needs to introduce three-body forces (TBF) which, in our approach are 
reduced to a density dependent two-body force by averaging over the position of 
the third particle \cite{bbb1,bbb2}. This allows to reproduce correctly the nuclear matter 
saturation point $\rho_0 \approx 0.17~\mathrm{fm}^{-3}$, $E/A \approx -16$ MeV, and gives
values of incompressibility and symmetry energy at saturation compatible
with those extracted from phenomenology \cite{myers}.

\begin{figure*}
\centering
\includegraphics[height=5cm,angle=270,clip]{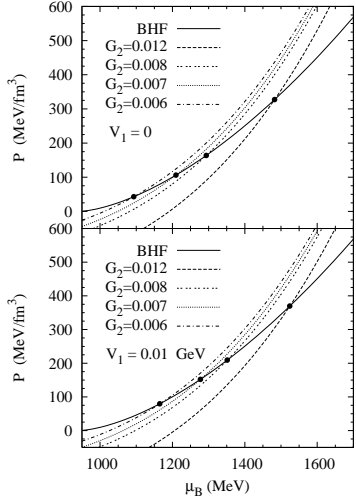}
\caption{Pressure as a function of the baryon chemical potential $\mu_B$. 
The full line represents the BHF calculations, 
and the dashed ones are the pressure derived in the FCM model 
for several values of the ($\mu_B$ independent) gluon condensate  $G_2$ 
and for two different choices of the parameter $V_1$ 
(upper and lower panel).
}
\label{fig2}     
\end{figure*}

Along with the BHF nonrelativistic EoS we consider its relativistic counterpart, i.e. the DBHF scheme \cite{fuchs}
where the Bonn A potential is used for the nucleon-nucleon interaction.
In the low density region ($\rho \rm < 0.3~ fm^{-3}$), both BHF (including TBF) and DBHF calculations are very similar, 
whereas at higher densities the DBHF is slightly stiffer  \cite{gabri}. 
The discrepancy between the nonrelativistic and relativistic calculation can be easily understood by recalling that the DBHF 
treatment is equivalent  to introducing in the nonrelativistic BHF  the TBF corresponding to the excitation of a nucleon-antinucleon pair, 
the so-called Z-diagram \cite{Z_diag}, which is repulsive at all densities. 
On the contrary, in the BHF treatment with Urbana TBF, both attractive and repulsive TBF are introduced
and therefore a softer EoS is expected.
\subsection{Quark Matter: the Field Correlator Method}
\label{sec-3}
The approach based on the FCM  provides a natural treatment
of the dynamics of confinement in terms of the
Color Electric   ($D^E$ and $D_1^E$)  and Color Magnetic ($D^H$ and $D_1^H$) 
Gaussian correlators, being the former one directly related to confinement,
so that its vanishing above the critical temperature implies deconfinement \cite{phrep}.
The extension of the FCM to finite temperature $T$ and chemical potential $\mu_q=0$ 
gives  analytical results in reasonable agreement with lattice data,
thus allowing  to describe correctly the deconfinement phase transition
\cite{sim4,sim22,sim6}.  In order to derive an EoS of the quark-gluon 
matter in the range of baryon density typical of the neutron star interiors, we 
have to extend  the FCM to finite values of the 
chemical potential \cite{sim4,sim22}. In this case, the quark pressure for a single 
flavour is simply given by \cite{sim4,sim22,sim6}

\begin{figure*}
\centering
\includegraphics[height=5cm,clip]{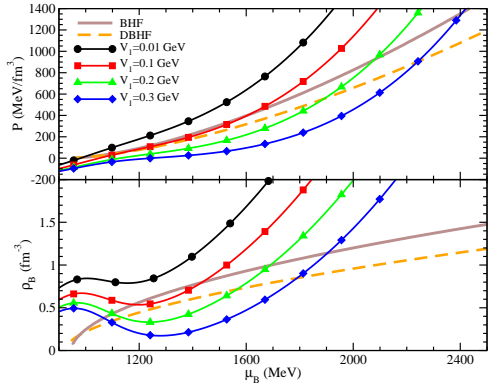}
\caption{
Pressure $P$ (upper panel) and baryon density $\rho_B$ (lower panel)
vs. $\mu_B$. The quark phase EoS is computed for various $V_1$ 
and with $G_2(\mu_B)$ given by the solid red line in figure~\ref{fig1}. 
For the hadronic phase, the solid brown line and  the dashed orange 
line respectively correspond to the BHF, and DBHF EoS.
}
\label{fig3}   
\end{figure*}

\begin{equation}\label{pquark}
P_q/T^4 = \frac{1}{\pi^2} \left[ \phi_\nu \left( \frac{\mu_q - V_1/2}{T} \right) +
\phi_\nu \left(-\frac{\mu_q + V_1/2} {T} \right ) \right ]
\end{equation}
where 
$\phi_\nu (a) = \int_0^\infty du  \left({u^4}/{\sqrt{u^2+\nu^2}} \right)
\left(\exp{ \left[ \sqrt{u^2 +\nu^2} - a \right]} + 1\right)^{-1} $
with   $\nu=m_q/T$,  and $V_1$ is the large distance static $\overline q~q$ potential which, in our analysis,
is treated as a free parameter.

The gluon contibution to the  pressure is 
\begin{equation}
\label{pglue}
P_g/T^4 = \frac{8}{3 \pi^2} \int_0^\infty  d\chi \chi^3
\frac{1}{\exp{\left(\chi + \frac{9 V_1}{8T} \right)} - 1}
\end{equation}
and the total pressure, that corresponds to the EoS in this phase, is 
\begin{equation}
\label{pqgp1}
P_{qg} = \sum_{j=u,d,s} P^j_{q} + P_g - \frac{(11-\frac{2}{3}N_f)}{32} \frac{G_2}{2}
\end{equation}
where $P^j_{q}$  and $P_g$ are respectively  given in Eq. (\ref{pquark}) and  (\ref{pglue}), and  $N_f$ is the flavour number.
The last term in Eq. (\ref{pqgp1})
corresponds to the difference of the vacuum energy density in the two phases
and $G_2$ is the gluon condensate whose numerical value, determined by the QCD sum rules 
at zero temperature and chemical potential, is known with large uncertainty: $G_2=0.012\pm 0.006~ \rm{GeV^4}$.

Therefore the EoS in  Eq.(\ref{pqgp1}) essentially depends on two parameters,
namely the quark-antiquark potential $V_1$ and the gluon condensate $G_2$.
In addition, at finite temperature 
and vanishing baryon density, a comparison with the recent available lattice calculations 
provides clear indications  about the specific values of these two parameters, and in particular 
their values at the critical temperature $T_c$. These estimates can be related to the corresponding 
values of the parameters at $T=\mu_B=0$
and, in particular, one finds   $ V_1(T=\mu_B=0) = 0.8 \div 0.9$ GeV \cite{bombaci,noi13}.
In our analysis, we are concerned about the dependence of $V_1$ and  $G_2$ on the  baryon 
chemical potential $\mu_B$  and, due to the absence of theoretical indications on $V_1$,
except for some lattice simulations that suggest no dependence of $V_1$ on $\mu_B$,
at least for  very small $\mu_B$  \cite{sim22,latmuf}, 
we treat $V_1$ as a free parameter with no dependence on $\mu_B$.

\begin{figure*}
\centering
\includegraphics[height=10cm,angle=270,clip]{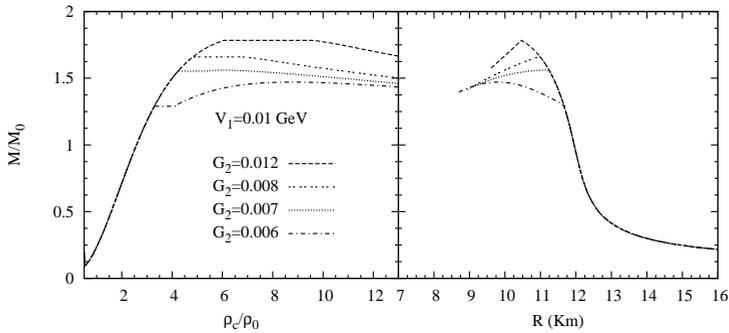}
\caption{The NS mass (in units of the solar mass) vs. the central baryon density (left panel) and the corresponding radius (right panel),
for $V_1=0.01$ GeV and some values of $\mu_B$ independent $G_2$. 
}
\label{fig4}       
\end{figure*}

On the other hand, there are indications about the dependence of the gluon condensate $G_2$ on $\mu_B$.
In particular, the QCD sum rules technique has been used to study
some hadronic properties within a nuclear matter environment at $T=0$ \cite{cohen,druc,balcasza}, and it has been found that the gluon 
condensate decreases linearly with the baryon density $\rho_B$, with small nonlinear corrections that can be neglected for  our purposes.
The corresponding curve is the dashed green line plotted in figure~\ref{fig1}.
According to this decreasing trend,  the  gluon condensate vanishes at some value of the baryon density and 
one expects that a transition to the deconfined state should  occur before reaching this point. 

A different analysis is presented in 
\cite{zit1,zit2} where $G_2(\mu_B) $ is computed in two-color ($N_c=2$) QCD, where many technical problems that affect the theory 
with $N_c=3$ are absent. In particular, the difference $f_{CS}(\mu)=G_2(\mu) - G_2(0)$, which  is computed from the energy momentum tensor 
of an effective chiral lagrangian, shows an initial decrease which, after reaching a minimum, is followed by a continuous growth,
and this trend is explained with the appearance of a weakly interacting gas of diquarks.
In \cite{zit1,zit2}, it is then suggested that $G_2(\mu_B) $ in full three-color ($N_c=3$) QCD has the same 
qualitative behavior  of the corresponding variable in two-color ($N_c=2$) QCD,
and therefore the plot of $f_{CS}(\mu)$, with the proper choice of the parameters for $N_c=3$ and fixing $G_2(\mu_B=0)=0.012$ GeV$^4$,  
is reported in figure~\ref{fig1} as the  dot-dashed blue curve.

Then, by following the first indication at low $\mu_B$ and the second one at larger $\mu_B$,
in our analysis we assumed a chemical potential dependent gluon condensate given by the solid red line in figure~\ref{fig1},
which starts at  $G_2(\mu_B=0)=0.012$ GeV$^4$ and approximates the dashed green curve at small $\mu_B$ and the 
dot-dashed blue one after the crossing of these two curves.
The analytic form of $G_2(\mu_B)$ is chosen to avoid unphysical features due to possible discontinuities in its derivatives.
Along with this particular choice of $G_2(\mu_B)$, we shall also display the results obtained with 
$\mu_B$ independent $G_2$.

\begin{figure*}
\centering
\includegraphics[height=4.7 cm,clip]{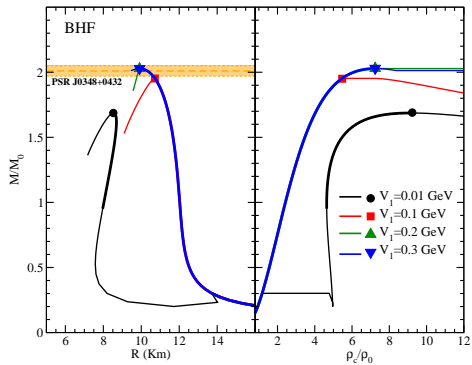}
\caption{The mass-radius (left panel) and the mass-central
density relation (right panel) for the BHF hadronic EoS and 
with $V_1$ and $G_2(\mu_B)$ as in figure~\ref{fig3}. 
The full symbols denote the value of the maximum mass. Stable 
configurations are displayed by thick lines, whereas thin lines 
indicate unstable configurations.
}
\label{fig5}     
\end{figure*}

\section{Hadron-quark phase transition and the corresponding NS masses}
\label{results}
The transiton between the hadron and quark phase is described by comparing 
the pressure of the two phases. We adopt the simple Maxwell construction, 
by assuming a first order hadron-quark phase transition and by imposing
thermal, chemical, and mechanical equilibrium between the two phases. 
This implies that the phase coexistence is determined by a crossing point 
in the pressure vs. chemical potential plot, as shown in figures~\ref{fig2} and \ref{fig3}.
In figure~\ref{fig2} the pressure $P$ is displayed as function of the baryon chemical potential 
$\mu_B$ in the nonrelativistic BHF case for the hadronic phase, and 
at very small  $V_1$ and  $G_2$ taken independent of $\mu_B$ for the quark phase.
Instead, figure~\ref{fig3} shows the pressure $P$ (upper panel) and the baryon density, 
$\rho_B=\partial P / \partial \mu_B$, (lower panel) vs.
$\mu_B$ for the hadronic EoS's (solid brown line for  BHF and dashed orange line for DBHF), 
whereas symbols are the results for quark matter EoS in the FCM and different choices of $ V_1$
and $\mu_B$ dependent $G_2(\mu_B)$, as introduced at the end of  Sect.~\ref{sec-3}.

We observe that,  both with increasing $G_2$ as in figure~\ref{fig2}
and  $V_1$ as in  figure~\ref{fig3}, the phase
transition point is shifted to larger chemical potentials which, in principle, indicates  
that the NS  possesses a thicker hadronic layer for larger values of these parameters.
In addition, while in figure~\ref{fig2} the derivative of the pressure in the quark phase 
always grows with $\mu_B$, 
in figure~\ref{fig3} there  is a region where the derivative of the pressure, $\rho_B$,
decreases (see lower panel). This behavior, as it is evident from a comparison of 
figures~\ref{fig2} and~\ref{fig3}, is due to the particular parametric form of 
the gluon condensate $G_2(\mu_B)$ shown in figure~\ref{fig1} and, as shown below,
it is the source of unstable neutron stars configurations.

The EoS is the fundamental input for solving the well-known hydrostatic equilibrium equations of 
Tolman, Oppenheimer, and Volkoff \cite{shapiro} for the pressure $P$ and the enclosed mass $m$
\begin{equation}
 \frac{dP(r)}{dr} = -\frac{Gm(r)\epsilon(r)}{r^2}
\left[ 1 + \frac{P(r)}{\epsilon(r)} \right]
\left[ 1 + \frac{4\pi r^3 P(r)}{ m(r)} \right]
\left[1-\frac{2Gm(r)}{r}\right]^{-1}
\label{tov1:eps}
\end{equation}
\begin{equation}
\frac{dm(r)}{dr} = 4\pi r^{2}\epsilon(r) 
\label{tov2:eps}
\end{equation}
being $\epsilon$ the total energy density ($G$ is the gravitational constant).
For a chosen central value of the energy density $\rho_c$, the numerical integration of
Eqs.~(\ref{tov1:eps}) and (\ref{tov2:eps}) provides the mass-radius relation.

\begin{figure*}
\centering
\includegraphics[height=4.7 cm,clip]{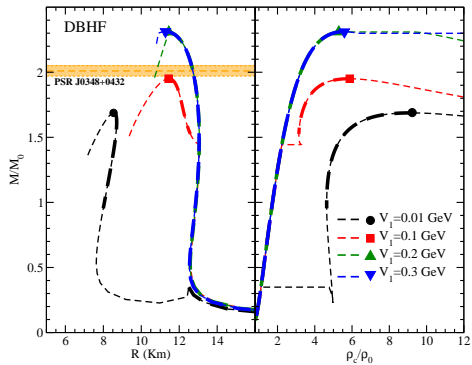}
\caption{Same as in figure~\ref{fig5} but with the DBHF EoS used for the
hadronic phase.
}
\label{fig6}     
\end{figure*}

We first consider the parametrization analyzed in the lower panel of 
figure~\ref{fig2} and we display in figure~\ref{fig4} the corresponding 
gravitational mass (in units of solar mass $\rm M_\odot = 2\times 10^{33}g$) 
as a function of the central baryon density  $\rho_c$,
normalized with respect to the saturation value  $\rho_0=0.17~{\rm fm}^{-3}$, (left panel)
and of the NS radius $\rm R$ (right panel). 
We observe that the maximum mass of the star is an increasing function of $G_2$
although its value is lower than the highest experimentally observed mass and therefore 
higher inputs for  $V_1$ are required.
However we recall that gravitationally stable NS configurations require ${\rm d M} /{\rm d} \rho_c > 0$ \cite{shapiro}
and then,  only those two curves with lower maximum mass in figure~\ref{fig4}
admit stable configurations with an inner quark matter core.

Finally we consider the more complex case with $\mu_B$ dependent gluon condensate,
presented in figure~\ref{fig3}. The corresponding NS masses are displayed in figures~\ref{fig5}  
and \ref{fig6},  respectively for the nonrelativistic BHF and relativistic DBHF hadronic EoS.
The orange band represents the recently observed  neutron star  PSR J0348+0432 \cite{maxpuls}.  
In this figures we have marked the gravitationally stable configurations by thick lines, whereas 
full symbols denote the maximum mass.  Unstable configurations are instead displayed by thin lines. 
Among the unstable configurations, we signal those characterized by increasing mass and decreasing 
central density, which are related to the nonmonotonic behavior of the derivative of the pressure 
in the quark phase in figure~\ref{fig3}, as anticipated  above.

For the hadronic BHF EoS in figure~\ref{fig5} the observational data require 
$V_1> 0.1$ GeV and, for $V_1> 0.2$ GeV the maximum mass does not grow beyond 2.03 solar masses \cite{noi13}.
In addition, in this range  of $V_1$, the maximum mass corresponds to the 
intersection of the hadronic and quark branches,
while stable configurations with a quark matter inner core are realized
only for  $V_1<0.095$ GeV, but with values of the maximum mass that are incompatible 
with the current observational data.

The stiffer structure of the hadronic 
DBHF  EoS yields larger NS masses as shown in figure~\ref{fig6}.
The qualitative picture is similar to that of figure~\ref{fig5}, but this time at $V_1 =0.1$ GeV we find 
stable NS with maximum mass compatible with observations and with a quark matter content.
We also find that for $V_1$ larger than  $\rm V_1 \approx 0.12$ GeV  the quark core disappears and for  $V_1> 0.2$ GeV 
the maximum mass stays at its  highest value of  2.31 solar masses.
\section{Conclusions}
\label{conclusions}
The study of the maximum mass of NS computed according to the FCM description of quark matter and 
to the nonrelativistic BHF or relativistic DBHF EoS of hadronic matter, provides some indications
concerning the parameters of the FCM when the results are compared with the recently
observed NS masses. Observational data require stiff EoS and  in fact the smoother hadronic BHF EoS
does not allow NS heavier than about 2 solar masses, which is still in agreement with the 
observational data, while the DBHF EoS reaches  2.31 solar masses. In any case such large mass
configurations can be reached only by increasing the two parameters $V_1$ and $G_2$ of the FCM,
because for small values of these parameters the typical maximum mass is  well below the observational
limit. However, while for small $V_1$ and $G_2$, stable configurations with an inner core of quark matter
are predicted, for those higher values of  $V_1$ and $G_2$ that predict sufficiently heavy masses,
the quark core disappears and the star has only hadronic content. The only small window
that predicts sufficiently heavy stable stars with a quark matter core is given by the DBHF EoS
combined with the FCM for $0.1~{\rm GeV} \lesssim V_1 \lesssim 0.12~{\rm GeV}$ and $G_2(\mu)$
parametrized as in Sect.~\ref{sec-3}.

Finally, it must be noticed that the relevant range of  $V_1$ found 
in this analysis ($0.1 \div 0.2$ GeV), is rather different from the one obtained from 
the FCM analysis of the transition at finite temperature ($0.8 \div 0.9$ GeV).
In \cite {noi13} it has been suggested that this difference could be an indication 
that, at finite density
and vanishing  temperature, the quark-antiquark potential $V_1$ is to be replaced by 
a quark-quark effective interaction  whose strength is about $V_1/4$.

\end{document}